\documentclass{aa}

\usepackage{graphics}
 
\thesaurus{06(08.02.01; 08.05.3)}

\newcommand {\belist}{\begin{list}{---}{\setlength{\rightmargin}%
{\leftmargin}}}

\newcounter{arabnum}
\newcommand{\befigcap}{\begin{list}{ {\bf Fig. \arabic{arabnum} } %
{ \usecounter{arabnum}} } }
\newcommand{\enfigcap}{\end{list}}

\newcommand{\bequo}{\begin{quotation}}
\newcommand{\enquo}{\end{quotation}}
\newcommand{\bverse}{\begin{verse}}
\newcommand{\everse}{\end{verse}}
\newcommand{\beit}{\begin{itemize}}
\newcommand{\enit}{\end{itemize}}
\newcommand{\been}{\begin{enumerate}}
\newcommand{\enen}{\end{enumerate}}
\newcommand{\ecen}{\end{center}}
\newcommand{\bcen}{\begin{center}}

\newcommand{\begeq}{\begin{equation}}
\newcommand{\eneq}{\end{equation}}
\newcommand{\befig}{\begin{figure}}
\newcommand{\enfig}{\end{figure}}

\newcommand{\ferrh}{\mbox{$\rm{[Fe/H]}$\ }}

\newcommand{\kmsec}{\mbox{${\rm \: km\:s^{-1}}$}}

\newcommand{\yrn}{\mbox{${\rm \:yr^{-1}}$}}

\newcommand{\uvby}{\mbox{\it uvby }}

\begin{document}

\title{Evidence for prolonged main sequence stellar evolution of F~stars  
in close binaries\thanks{Based on the data from the Hipparcos astrometry 
satellite (European Space Agency)}
}
 
\titlerunning{Main sequence stellar evolution in close binaries} 

\author{A. A. Suchkov \inst{}}

\institute{Space Telescope Science Institute\thanks{
Operated by AURA Inc., under contract with NASA, Baltimore, MD 21218, USA}}

\date{Received 16 November 2000 / Accepted 23 January 2001}

\maketitle

\begin{abstract}
Binary F~stars exhibit large brightness anomaly, which is defined here as the 
difference between the  absolute magnitude from the \uvby photometry and  the 
actual absolute magnitude of the star. We have found that the  anomaly 
inversely correlates with the binary components separation. There is evidence 
that the correlation reflects actual population differences between close 
and wide  binary pairs, in which case it indicates that the anomaly is somehow 
associated with the interaction of binary's components. The anomaly has also 
been found to correlate with both kinematics and metallicity. The sense 
of the correlations implies that the anomaly increases as the star evolves, 
suggesting a peculiar evolution of a primary F~star  in a tight binary pair. 
This conclusion has further been supported by the study of the age--velocity
relation (AVR) of F~stars that are cataloged in the {\it HIPPARCOS}  
as single. Among these stars, those with brightness anomaly were previously 
shown to be most likely unidentified close binaries. We have found that the 
AVR of these binary  candidates is different from that  of the ``truly single'' 
F~stars. The discrepancy between the two AVRs indicates that the putative 
binaries are, on average, older than similar normal single F~stars at the same 
effective temperature and luminosity, which is consistent with the inferred 
peculiar evolution in close binaries. It appears that this peculiarity is 
caused by the impact of the components interaction in a tight pair on stellar 
evolution, which results in the prolonged main sequence lifetime of the primary 
F~star. 

\end{abstract}
\section{Introduction 
\label{sec-intro}  
} 
Many of the F stars cataloged in the {\it HIPPARCOS}  as single
were recently argued to be in fact unidentified binaries 
(Suchkov \&\ McMaster 1999). 
The criterion used to isolate binary candidates involves the difference
between the absolute magnitude $M_V$ based on the {\it HIPPARCOS}  parallax
and the absolute magnitude $M_{c_0}$ derived 
from the \uvby luminosity index $\Delta c_0$, i.e.,   
$\Delta M_{c_0} = M_{c_0} - M_V$. 
For the best 
known nearby single F~stars ($d < 25$~pc), the two magnitudes coincide within 
$\pm 0.15$~mag. However, most of the unresolved binary stars  are anomalously
bright for their \uvby colors, i.e., their absolute magnitude, $M_V$, is 
substantially brighter than the magnitude derived from $\Delta_{c_0}$: 
$\Delta M_{c_0} > 0.15$ 
(for simplicity, we will often be referring to $\Delta M_{c_0}$  as
brightness anomaly and consider a star as 
anomalously bright  if $\Delta M_{c_0} > 0.15$).   
Unlike the nearby single stars, many of the distant single F stars, 
($d > 25$~pc) turned out to be anomalously bright as well, similar to the
known binary F~stars. The analysis showed that in this case, 
the brightness anomaly is likely associated with the presence of undetected 
companions (Suchkov \&\ McMaster 1999). These binary candidates were called 
C~binaries (the follow-up radial-velocity survey of northern 
C~binaries with $\Delta M_{c_0} \geq 0.5$ and  $d < 80$~pc conducted by 
Griffin \&\ Suchkov 2001 has already confirmed that more than 
$35\%$ of the sample stars actually {\it are} binaries; orbits
for many of these binaries have been determined). 

For unresolved binaries, $M_{c_0}$ is derived from the
color indices originating from a double star combined spectrum. Because of 
that $\Delta M_{c_0}$  underestimates the luminosity of that star if its
components are comparably bright, by up to $-2.5\log 2 \approx -0.75$~mag 
in the case of identical components (this obviously imposes an upper limit
of 0.75  on $\Delta M_{c_0}$).  However, this seems to be not the only, 
and perhaps not even the main effect leading to the discrepancy
between $M_{c_0}$ and $M_V$. For one thing,  a substantial fraction of
binary  stars have brightness anomaly well above the indicated upper limit 
of 0.75~mag. The stars with $\Delta M_{c_0} > 0.75$, which cannot be 
explained in terms of the  combined flux of the unresolved binary components, 
indicate that along with the combined luminosity something else may impact 
the absolute magnitude of unresolved binary  stars.

More evidence favoring the existence of an additional source 
for brightness anomaly has come
from the study of the relationship between $\Delta M_{c_0}$ and stellar age
(Suchkov 2000). It turns out  
that in terms of kinematics and metallicity, anomalously bright binary
candidates, C binaries, are, on average, substantially  older than 
the ``truly single''  F stars, i.e., the stars with normal brightness.
If the anomaly were caused entirely by the luminosity contribution from the
undetected secondary, one would have to conclude that binaries  with 
the components having about the same mass live longer or
tend to form earlier in the galactic evolution  than binaries with
disparate components. But such a scenario does not seem very likely.
A more plausible interpretation of the relationship between 
$\Delta M_{c_0}$   and age might be that at least some part of the star's 
brightness anomaly is associated with a peculiar stellar evolution in 
the binary system, so that the primary  F~star in that binary stay within 
the  main sequence longer than a normal single  F~star.
Given the obviously important implications of such  a possibility, we
have examined a number of relationships between 
$\Delta M_{c_0}$  and other parameters of F~stars in order to get a deeper 
insight into the nature of brightness anomaly.
\section{Data
\label{sec-data}
}
The star sample used in this study originates from the list of 
$\sim 10,000$ {\it HIPPARCOS} stars that have measured   \uvby colors,
with effective temperatures corresponding to the spectral range occupied
by F and early G stars (Suchkov \&\ McMaster 1999).
The \uvby data are from Hauck \&\ Mermilliod (1998). 
We have computed tangential velocities of these stars from the {\it HIPPARCOS} 
parallaxes and proper motions. The {\it HIPPARCOS}  parallax has also been 
utilized to derive absolute magnitude $M_V$ from the Johnson $V$ magnitude 
and absolute magnitude $M_{Hp}$ from the {\it HIPPARCOS}  $Hp$~magnitude.  
Absolute magnitude $M_{c_0}$ has been calculated from the \uvby data with
the algorithm published by Moon (1985; see also Moon \&\ Dworetsky 1985); 
these calculations involve the dereddened \uvby luminosity index, 
$\Delta c_0$, along with the dereddened metallicity and temperature indices.  
The algorithm in Moon (1985) has also been used to compute
standard $\beta$ from the blanketing-corrected color index $(b-y)_{cor}$,
from which effective temperature has been obtained.

Metallicity has been calculated from the dereddened \uvby color indices,
utilizing the calibration given by Schuster \&\ Nissen (1989) for F stars.  
This calibration is somewhat different from that in Carlberg et al. (1985), 
which was used in Suchkov (2000) and Suchkov \&\ McMaster (1999).
This difference, however,  has been found not to impact the results of 
both the previous and present studies. 
The Schuster \&\ Nissen (1989) calibration is commonly utilized in current 
studies, so we will use it in this paper.

Age estimates are based on isochrone fitting in the 
$\log T_e - M_V$  diagram using Yale 1996 isochrones (Demarque et al. 
1996\footnote{Available at http://shemesh.gsfc.nasa.gov/astronomy.html}). 
Age has been used only in terms of mean values 
obtained by averaging over large groups of stars, and  only in differential 
analysis involving relative ages. Therefore, age accuracy, both in terms of 
random and systematic errors, is not of  much concern for the present study.

The sample has been constrained  as follows:
$0.22 \leq (b-y) \leq 0.39 $;
$5800 \leq T_e \leq 7500 $,~K;
$-0.6 \leq \ferrh \leq 0.5$; 
$0.02 \leq e_{c1}$ ($e_{c1}$ is the mean error on $c_1$ for an individual star
 as given in Hauck \&\ Mermilliod 1998; the sample mean for $e_{c1}$ is 
$\sim 10$ times smaller);
$0.015\leq e_{m1}$ ($e_{m1}$ is the mean error on $m_1$ for an individual star);
$3 \geq e_{\pi}$,~mas  (error on parallax);
$3 \geq e_{\mu_\alpha}$,~mas~\yrn (error on proper motion in RA);
$3 \geq e_{\mu_\delta}$,~mas~\yrn (error on proper motion in DEC).  
\section{Results and discussion}
\subsection{Brightness anomaly distribution}
\begin{figure}
\resizebox{\hsize}{!}{\includegraphics{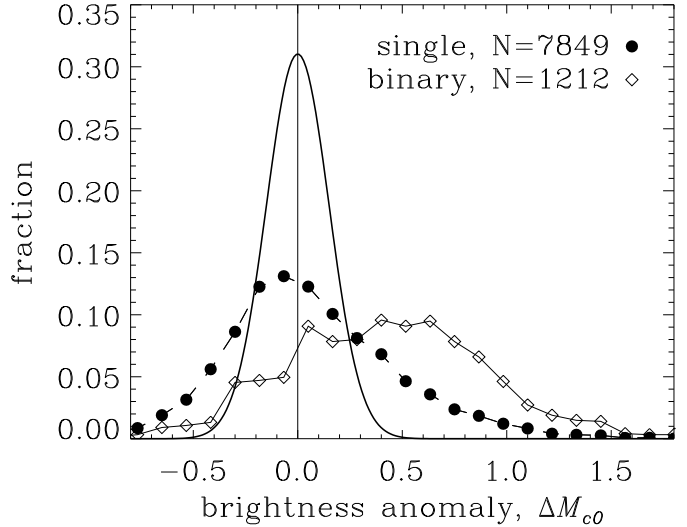}}
\caption{Normalized brightness anomaly distribution for the  unresolved 
binaries  and the stars cataloged in the {\it HIPPARCOS} as single.
The gaussian is for the parameters derived from the sample of 
single F~stars within  25~pc.
}
\label{H2541F1}
\end{figure}

Brightness anomaly is very  conspicuous in unresolved binary stars 
(Fig.~\ref{H2541F1}). As shown in Suchkov \&\ McMaster (1999),
the distribution of $\Delta M_{c_0}$  for the best studied single stars
within 25~pc is centered at $\Delta M_{c_0} = 0$  and is rather narrow, 
with standard deviation of only 0.15~mag.  The corresponding gaussian 
($\sigma_{\Delta M_{c_0}} = 0.15$) is shown in Fig.~\ref{H2541F1}.
For the unresolved binaries, the maximum of the $\Delta M_{c_0}$  distribution 
is not only offset from zero by $\sim 0.5$~mag but is also much broader. 
Within 200~pc, the stars cataloged  in the {\it HIPPARCOS} as single have a 
broad distribution as well, but with the maximum only slightly shifted  
($\sim -0.1$~mag) from that of the single stars within 25~pc. 

Note that the left tail of the distribution, albeit resembling  a segment of a 
gaussian, is far above the level suggested by $\sigma_{\Delta M_{c_0}} = 0.15$,
meaning  that it does probably not represent the error distribution for 
$\Delta M_{c_0}$. Therefore,  the  population of anomalously faint stars, 
$\Delta M_{c_0} < 0.15$, should be as real as that of anomalously bright stars.
As seen below, its distinctive identity is reflected in lowest velocities 
and highest metallicities  among the F~stars,  which makes it the youngest
group of F~stars. Therefore, we conclude that in general
the distributions of both binary  and single stars in 
Fig.~\ref{H2541F1} are most likely dominated by some stellar 
population physics rather than random or systematic errors.

The most striking feature in Fig.~\ref{H2541F1} is the
positive brightness anomaly of binary stars. Therefore, in the rest 
of the paper, we will focus primarily on the nature 
of anomalously bright binary stars. 

\subsection{Correlation between $\Delta M_{c_0}$  and binary components
separation}

If the combined flux of unresolved binary components is not the
only reason for the enhanced brightness of a binary star, and
there is a noticeable contribution from a physical anomaly in the 
binary's primary star  caused by its interaction with the companion, one may 
expect the brightness anomaly to depend on binary components separation, 
because the effect of the interaction,  whatever it is, is probably stronger 
in tighter pairs.  
We have tested this hypothesis by correlating brightness anomaly and projected 
separation for two very different, non-overlapping samples of binary F~stars
within 125~pc, for which the {\it HIPPARCOS}  provides angular separation, 
$\rho$. Only the stars with $\Delta M_{c_0}$  in the range $-0.5$ to~1.0 
have been included, so that the potentially less reliable extreme values 
of $\Delta M_{c_0}$  do not impact the results. 
The chosen distance limit rejects the most  distant 
stars, thus ensuring better data quality for the remainder of the sample;
at the same time it retains enough stars for statistical tests we are 
interested in.

\begin{figure}
\resizebox{\hsize}{!}{\includegraphics{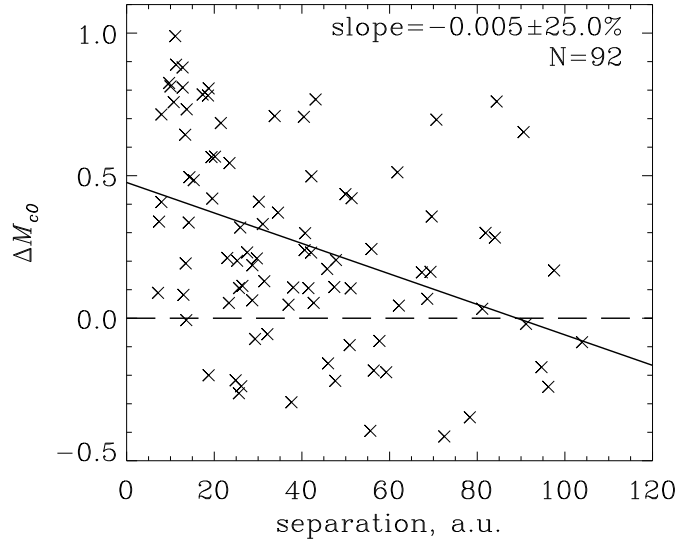}}
\caption{Correlation between brightness anomaly
and binary components  separation for a distance limited
sample ($d \leq 125$~pc) of binary  F~stars 
discovered by {\it HIPPARCOS}. The sample stars have  
 angular separation  within 1~arcsec. 
}
\label{H2541F2} 
\end{figure}
 
\begin{figure}
\resizebox{\hsize}{!}{\includegraphics{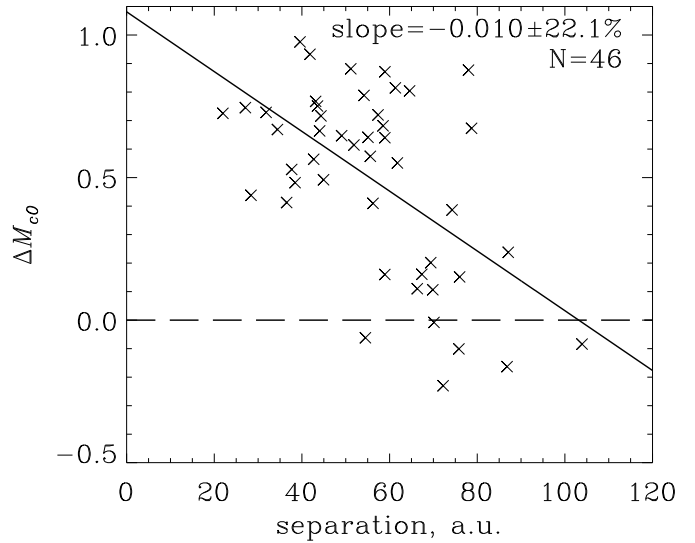}}
\caption{Correlation between brightness anomaly
and binary components  separation for unresolved pre-{\it HIPPARCOS}  
binary  F~stars within 125~pc with angular separation $\rho \leq 1.0$~arcsec. 
}
\label{H2541F3} 
\end{figure}

The stars in the first  sample are binaries discovered by {\it HIPPARCOS}
(see Mignard et al. 1992, S\"oderhjelm et al. 1992 for the procedures
to detect and measure binary stars on the basis of {\it HIPPARCOS}  
observations).  Only the  binaries with angular separation 
$\rho \leq 1$~arcsec have been included; this constraint
rejects only a very small fraction of the  {\it HIPPARCOS}  binaries 
and is not of great significance for the test below.
The sample has also been constrained with respect to angular separation error,
$\epsilon_{\rho}$, to include only the stars with $\epsilon_{\rho}/\rho < 0.1$.

The second sample comprises unresolved binaries known prior to {\it HIPPARCOS}. 
With respect to angular separations, it has been constrained similar
to the sample of the {\it HIPPARCOS}  binaries,
$\rho \leq 1$~arcsec. The pre-{\it HIPPARCOS}  binaries are much more 
numerous,  which allows us to impose a more stringent constraint on the 
angular separation accuracy: $\epsilon_{\rho}/\rho  < 0.01$. 
Additionally, this sample has been constrained to include only hot stars, 
$6500 - 7500$~K. This should improve chances to detect any separation-dependent
evolutionary effects in $\Delta M_{c_0}$ just because the fraction of 
evolved stars increases toward higher temperatures (at cooler temperatures, 
brightness anomaly in the sample stars may be dominated by the 
separation-independent effect of the combined flux of binary's unresolved 
components).

Projected separation for both samples has been derived from the parallax, $\pi$,
and the angular separation, $\rho$, given in the {\it HIPPARCOS} catalogue.

The stars of the two samples  are presented 
in Fig.~\ref{H2541F2} in  Fig.~\ref{H2541F3}.
As one can see, they do indeed show a correlation between 
separation and brightness anomaly, with the slope of the linear regression 
being non-zero well above the error in both cases (the slope error is given
as a percent of the slope value). 

It is to be noted that the correlation
remains the same when absolute magnitude $M_V$ in the definition of 
$\Delta M_{c_0}$  is replaced with absolute magnitude $M_{Hp}$ obtained
from the {\it HIPPARCOS}  $Hp$ magnitude (for the details of 
{\it HIPPARCOS}  photometry see Mignard, Froschle, \&\ Falin 1992;
Evans et al. 1992). The corresponding regression slopes are:
$-0.005 \pm 25.2$\% for the {\it HIPPARCOS}  binaries, and 
$-0.011 \pm 21.6$\% for the pre-{\it HIPPARCOS}  binaries.

The correlation in Fig.~\ref{H2541F2} and Fig.~\ref{H2541F3}
reveals that $\Delta M_{c_0}$  is, on average, larger at 
smaller component separations. This suggests that brightness anomaly is 
somehow associated with the components interaction in a tight binary pair. 
The question is then: what is the physics behind this anomaly?

\subsection{Relationship between $\Delta M_{c_0}$  and stellar kinematics }
To answer the above question, we have explored whether
there is any dependence in $\Delta M_{c_0}$ on age, because the effects
caused by interaction in a close binary may accumulate with time.  
First of all, we have correlated $\Delta M_{c_0}$  and tangential 
velocity, employing the fact that the statistics from stellar kinematics, such 
as mean tangential velocity and velocity dispersion used here, are
age-sensitive. 

\begin{figure}
\resizebox{\hsize}{!}{\includegraphics{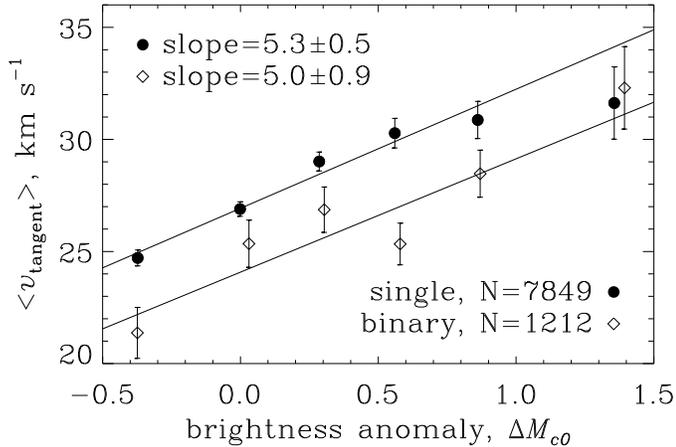}}
\caption{Mean tangential velocity versus $\Delta M_{c_0}$  for 
the unresolved binary F stars  and F~stars  cataloged in the 
{\it HIPPARCOS}  as single. 
Weighted linear regression and its slope are shown for both groups of stars.
}
\label{H2541F4} 
\end{figure}

As seen in Fig.~\ref{H2541F4}, there is indeed  a significant 
correlation between  $<\!\!v_{\rm tangent}\!\!>$  and $\Delta M_{c_0}$.
It implies that   $\Delta M_{c_0}$ is larger, on average, at older ages.
Given that the {\it HIPPARCOS}  single stars with large $\Delta M_{Vc_0}$  
are probable close binaries, we conclude  that this result is consistent
with increase in brightness anomaly for a tight binary pair 
as the primary star evolves.

Fig.~\ref{H2541F4} also reveals that the kinematics 
of known binary stars is ``younger'' than that of the single stars at the same 
$\Delta M_{c_0}$. This reflects the generally younger age of the population 
of the known binary F~stars, which  was discussed in  Suchkov (2000). 
\subsection{Relationship between $\Delta M_{c_0}$  and metallicity }
\begin{figure}
\resizebox{\hsize}{!}{\includegraphics{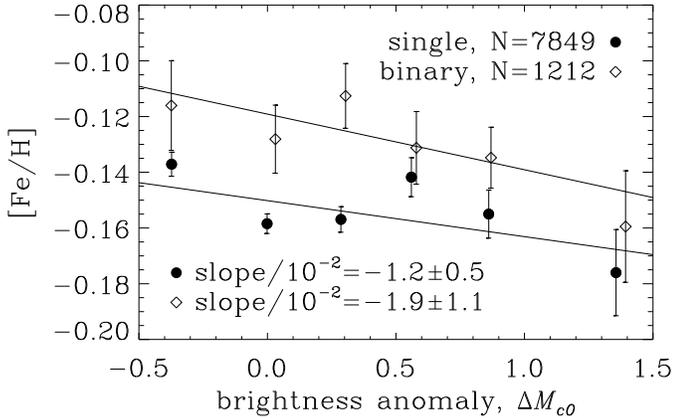}}
\caption{Metallicity versus $\Delta M_{c_0}$ for the unresolved binary 
F~stars and F~stars  cataloged in the {\it HIPPARCOS}  as single. 
Weighted linear regression and its slope are shown for both 
groups of stars.
}
\label{H2541F5}
\end{figure}

Mean metallicity is another age-dependent statistics that can be used to 
check if there is any relationship between $\Delta M_{c_0}$  and age.
The correlation between \ferrh\ and $\Delta M_{c_0}$  is shown in 
Fig.~\ref{H2541F5} for the same groups of stars  as in 
Fig.~\ref{H2541F4}. Here, the linear regression slope is  non-zero
within more than one sigma for both the binary and single star samples,
indicating that $\Delta M_{c_0}$  is larger at lower metallicities. Since 
lower metallicity is indicative of older age, the correlation in 
Fig.~\ref{H2541F5} is consistent with the inference from the preceding 
subsection that the stars with larger $\Delta M_{c_0}$  are, on average,
older.

The correlation in Fig.~\ref{H2541F5} could have been expected 
on the basis of the correlation in Fig.~\ref{H2541F4} because metallicity
is known to correlate with kinematics. At the same time this correlation 
is obviously  weaker than that in 
Fig.~\ref{H2541F4}. Since metallicity is also known to have a large spread 
at any given age,  a possible interpretation of weaker correlation
may be that metallicity is  coupled with age more loosely than kinematics. 

The weaker correlation in Fig.~\ref{H2541F5}, whatever its reason, 
argues against the possibility  that
the correlation in Fig.~\ref{H2541F4} results from a metallicity-dependent 
bias in estimates of $M_{c_0}$, hence $\Delta M_{c_0}$. If such a bias did
exist (say, due to flaws in the algorithm that computes $M_{c_0}$) and cause
all the correlation in Fig.~\ref{H2541F5}, one would have expected it to 
translate into a weaker rather than stronger correlation between
$\Delta M_{c_0}$  and kinematics in  Fig.~\ref{H2541F4}. 
It is to be added that the aforementioned hypothetical  bias seems unlikely 
anyway, because the available evidence, including the results from the binary 
components separation data and the age--velocity relation (see below), 
argues against any role of this kind of systematic errors in brightness 
anomaly.  

Thus, the  interpretation of both Fig.~\ref{H2541F4} and 
Fig.~\ref{H2541F5} involving the dependence of brightness anomaly on age
appears to be reasonably justified.

\subsection{Age--velocity relation: discrepancy between normal and anomalously
bright stars}
So far, we have been referring to brightness anomaly in the sense 
that the absolute magnitude of a star with this anomaly is brighter than 
that of a normal single star with the same \uvby colors. But it would be 
equally correct to say that the star has  the \uvby colors 
that are anomalous for a normal  star at a given effective temperature and 
luminosity. The anomaly in this case refers to the color index~$c_0$.

For the main sequence F~stars, $c_0$ provides a measure of 
Balmer discontinuity (see Crawford 1975 and references therein). 
In these stars, smaller Balmer discontinuity at a given effective temperature 
(hence  smaller $c_0$) means a denser stellar atmosphere,  i.e.,  larger 
surface gravity. In this way, $c_0$ actually  measures  stellar surface gravity
(at a given $T_e$), through which it is related to the star's luminosity. 
Anomalous $c_0$ in an F~star then  means anomalous surface gravity, $\log g$. 
Specifically, the anomaly resulting in a positive $\Delta M_{c_0}$  corresponds
to the surface gravity that is too high for a given effective temperature, 
$T_e$, and luminosity, $L$. Consequently, the stellar mass of such a star is 
larger than the mass of a normal single star at the same $T_e$ and $L$. 
(Alternatively, one can say that the  luminosity of an anomalously  bright 
star is too high for a given $T_e$ and $\log g$). This obviously requires
that stellar evolution of a normal star is different from that of an 
anomalously bright star. So, if $c_0$ of the  anomalous stars in our sample
is dominated by the effect of enhanced surface gravity, these stars  can be 
predicted to evolve in a non-standard way.
\begin{figure}
\resizebox{\hsize}{!}{\includegraphics{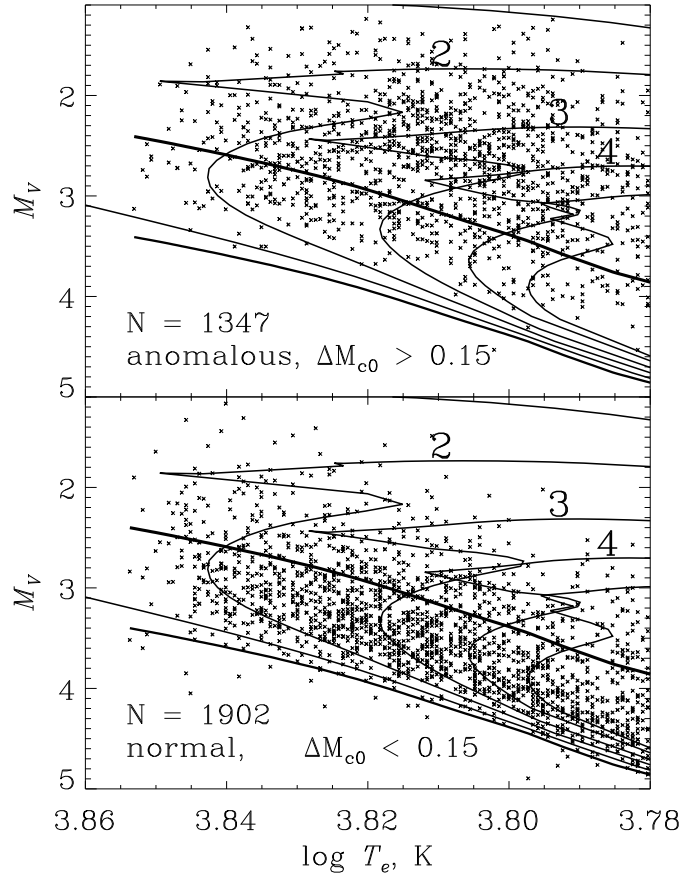}}
\caption{$\log T_e - M_V$  diagram of the {\it HIPPARCOS} ``single'' F~stars.
Upper panel: anomalously bright stars, presumably dominated by unidentified
close binaries (C binaries). Lower panel: stars with mostly normal brightness,
presumably comprising truly single stars as well as unevolved close binaries
with disparate components.
The isochrones are labeled by their age in Gyr.
The level of 1.0 mag above ZAMS is shown (see text for details).
}
\label{H2541F6} 
\end{figure}

We can verify the latter prediction as follows. Let us take the sample
of the {\it HIPPARCOS} ``single'' stars and split
it into two subsamples: the first one including 
only anomalously bright stars, $\Delta M_{c_0} > 0.15$ (C~binaries), 
and the second subsample that includes the stars  with 
$\Delta M_{c_0} < 0.15$; for simplicity, the latter stars will be referred 
to as ``truly single'', although  we expect a fraction of them to be
unevolved and/or wide binaries with disparate components.
Now, we can derive and compare the age--velocity relations (AVR) for these
two group of stars.  If the same stellar evolution model (meaning the
same set of isochrones used to derive age) is applicable to all stars,
we should obviously get the same AVR for both groups, simply  because
stellar kinematics and stellar evolution are entirely unrelated things.
If this proves not to be the case, and the age--velocity relations are
discrepant, the two groups of stars evolve probably differently.

\begin{figure}
\resizebox{\hsize}{!}{\includegraphics{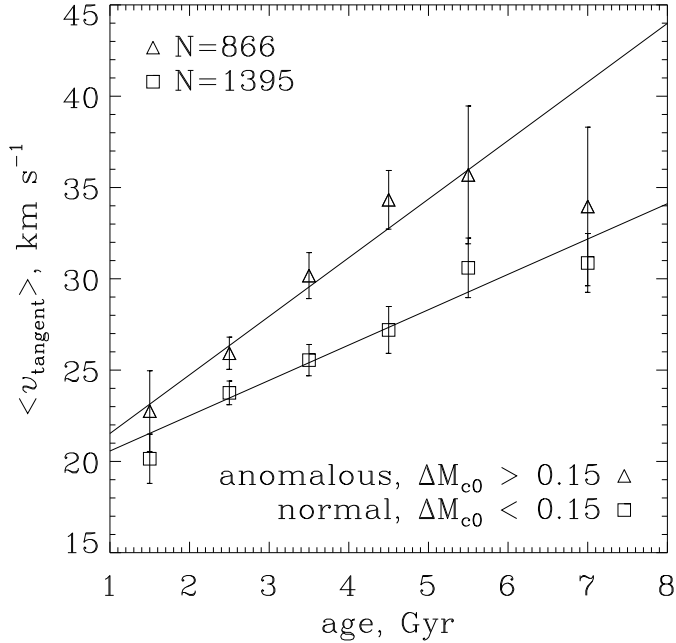}}
\caption{Age--velocity relation (AVR) for the normal and anomalously bright 
F~stars (see text for details).
}
\label{H2541F7} 
\end{figure}

We have derived  ages for both groups of stars based on isochrones from 
Demarque et al. (1996). 
The isochrones as well as the stars used to calculate the age--velocity 
relations are shown in Fig.~\ref{H2541F6}.  
The stars have been selected in a narrow metallicity range, 
$-0.3 < \ferrh < -0.1$, that matches the chemical 
composition of the isochrones, $Z = 1.0 \times 10^{-2}, \;\; Y=0.27$.
Additionally they have been constrained to the distance range $d \leq 125$~pc.
They have been grouped into 1 to 2~Gyr age bins, for which mean tangential 
velocity has been calculated. Because of large age errors for the young 
stars near the ZAMS, ages younger 1~Gyr have not been considered. 
The resulting age--velocity relations for the age range from 1~to~8~Gyr
are given in  Fig.~\ref{H2541F7}. 

\begin{table}
 
\caption{Kinematics of normal (``truly single'') and anomalously bright
(C~binaries ) F~stars above and below  the line $M_{ZAMS} - M_V = 1.0$ 
shown in Fig.~\ref{H2541F6}. 
}
\begin{tabular}{lcccc}\\ \hline \hline
 
brightness \ \ \ \ \ \ \ \
& $\Delta M_{c_0}$		
& $N$ 
& $<\!{v_{\rm tangent}\!>}$ 
& $\sigma_{\rm tangent}$ \\

{}
& (mag)   
&  {}  
&   (\kmsec)    
&   (\kmsec)    \\ \hline
 
{} & {} & \multicolumn{3}{c}{ $M_{ZAMS} - M_V > 1.0 $~mag } \\

normal \dotfill
&$< 0.15$		
& 1226
&  26.4 $\!\pm\!$ 0.5
&  16.5 $\!\pm\!$ 0.3 \\

anomalous \dotfill
&$> 0.15$ 		
& 2550
&  30.2 $\!\pm\!$ 0.4
&  18.2 $\!\pm\!$ 0.3 \\

{} & {} & \multicolumn{3}{c}{ $M_{ZAMS} - M_V < 1.0 $~mag } \\

normal \dotfill
&$< 0.15$ 		
& 3416
&  26.0 $\!\pm\!$ 0.3
&  16.6 $\!\pm\!$ 0.2 \\

anomalous \dotfill
&$> 0.15$ 		
& 876
&  29.2 $\!\pm\!$ 0.7
&  20.9 $\!\pm\!$ 0.5 \\

\end{tabular}

\end{table}

Fig.~\ref{H2541F7} reveals  a conspicuous discrepancy between the two 
age-velocity relations, which will be referred to as the AVR~discrepancy.
The possibility that kinematic differences between the normal and anomalously
bright stars arise from different distance sampling (the  latter stars
are, on average, more distant) was checked and ruled out in Suchkov (2000).
We have also looked into the possibility that the bias associated with 
the brighter, on average, absolute magnitudes of anomalously bright stars 
(see Fig.~\ref{H2541F6})  plays a role in the  discrepancy (for example, 
this may happen if the isochrones underestimate the age of far evolved stars).
To this end,  we have compared the kinematics of the normal and anomalously 
bright stars with $-0.6 < \ferrh < 0.3$ in the two regions of the 
$\log T_e - M_V$  diagram, above and below the line corresponding to 1.0~mag 
above the ZAMS shown in Fig.~\ref{H2541F6}.
As seen in Table~1, anomalously bright stars close to the ZAMS have velocities 
higher, on average, than those of the normal stars far above the ZAMS. This 
rules out the indicated bias as the cause of the AVR discrepancy. Therefore, 
the discrepancy must be real, reflecting some inconsistency in the derived ages. 
Assuming that the isochrones in Fig.~\ref{H2541F6} adequately represent 
the age of the normal stars (truly single), we must then conclude that the age 
of the stars with large brightness anomaly (C binaries) is underestimated. 
The amount of the AVR discrepancy suggests that the underestimation is, 
on average, as large as 2~Gyr or more.  Thus, anomalously bright stars are 
consistent with being actually older than normal stars at the same positions 
in the $\log T_e - M_V$  diagram. 

There are at least two reasons why isochrone fitting  could be inadequate 
for anomalously  bright stars and result in incorrect ages: (i) combined
flux of unidentified binaries with comparably bright components, and
(ii) non-standard stellar evolution.
 
If anomalously bright stars are unidentified
binaries with comparably bright components, isochrone fitting
results in either overestimated or underestimated age, depending on 
the position of the star in the $\log T_e - M_V$  diagram and the contribution
of the secondary into the star's $V$~magnitude. 
For the  stars cataloged in the  {\it HIPPARCOS} as single, 
age would be, on average, underestimated (Suchkov \&\ McMaster 1999). 

The following, however, argues against the hypothesis that such binaries
dominate our sample of anomalously bright stars.

First, the  sample stars are much older (on average) 
than the known unresolved binaries (Suchkov, 2000; see also 
Fig.~\ref{H2541F4} and Fig.~\ref{H2541F5}). This means that
most of these stars are different from normal binaries. Even if they 
are double systems, in most cases their secondary component is hardly 
a regular star whose  emission is comparable to that of the primary.

Second, as mentioned before, even among known binary stars a substantial 
fraction have brightness anomaly exceeding the maximum value of 0.75~mag, 
hence it cannot be attributed to the combined flux of the binary 's two 
components only (see Fig.~\ref{H2541F1}). Statistically, the excess 
is quite significant, so there is little doubt that in many cases the cause 
of brightness anomaly is different from the light contribution from the 
star's unresolved companion.

So, it seems unlikely that the AVR discrepancy in Fig.~\ref{H2541F7} 
is due only to binaries with comparably bright components.  

The above isochrone fitting can also be inadequate 
if for some reasons the evolution of a star is different from
the normal evolution of a single star. 

In such a case the  
position of a star in the $\log T_e - M_V$  diagram would correspond to 
an age different from the one predicted by standard
isochrones.  The age--velocity relation based on ages from the standard 
isochrones  would be incorrect for these stars, and this may be the reason for
the AVR discrepancy revealed by Fig.~\ref{H2541F7}.

As discussed below, anomalous stellar  evolution, unlike the effect
of combined flux, offers a way to explain the relationship between age 
and  brightness anomaly; in particular, it explains why C~binaries are,
on average, older than the truly single stars. Also anomalous evolution
appears  to be in line with the recent developments in studies of eclipsing 
binaries. This makes it a stronger candidate for being the
dominant source of the AVR~discrepancy in Fig.~\ref{H2541F7}.
The combined evidence discussed in this paper suggests then that 
anomalous stellar evolution occurs apparently  in tight binary pairs
where the evolution of the primary star is affected by the presence
of the companion.
\section{Discussion and summary}
The older ages of anomalously bright stars (Suchkov 2000) seem to indicate 
that these stars stay within the main sequence longer than normal single stars. 
Such a prolonged main 
sequence evolution is known to occur in stars with enhanced central mixing 
in the convective core, and is apparently quite ubiquitous among close
binary systems at stellar masses slightly above solar (see, e.g., Chiosi 1999,
Andersen \&\ Nordstr\"om 1999, Gimen\'ez et al. 1999).
This suggests that  anomalous brightness discussed in this paper
reflects non-standard stellar evolution that involves enhanced core mixing. 
Extreme manifestation of brightness anomaly among known unresolved binary 
stars seems to indicate that the anomaly is somehow related to binarity.
Given that $\Delta M_{c_0}$  correlates with binary components separation,
one may infer that interaction in a tight binary pair somehow induces 
extra mixing in the convective core of the binary's primary component,
additional to whatever mixing occurs  in a similar single star.
Of course, there are other  mechanisms that may impact stellar evolution
in a  tight binary pair. One may recall, for example, modified convection 
in the outer convective envelope of a star in a tight binary pair  or 
mass loss/transfer  in very close  binary systems. These mechanisms were 
discussed in the literature as related to the existence of blue stragglers and 
other unusual stars in stellar clusters  (Pols \&\ Marinus 1994), apparent age 
discrepancy between the components in low mass binaries (Clausen et al. 1999), 
young age of the known unresolved binary stars  (Suchkov 2000), etc. However,
enhanced convective core mixing seems to be the only mechanism capable of
extending the  main sequence evolution by a significant amount. 

It is to be noted that despite the potentially longer than normal lifetime,
the actual average main sequence lifetime of the  known unresolved binary stars
is  shorter than that of the single stars (Suchkov 2000). As hypothesized in the
latter paper, this may be related to violent interaction of the binary's
stellar components, which effectively removes the binary's primary in the
color--magnitude diagram from the region occupied by the  main sequence 
F~stars; mechanisms of such an interaction were considered, for instance, 
in Pols \&\ Marinus (1994). On the other hand, C~binaries were found to be
indeed significantly older, on average, than the truly single stars, possibly
indicating that many of C~binaries have substellar secondaries incapable
of causing too much damage to the primary F~star.


Unidentified close binaries may represent a large fraction of the 
local population of F~stars. Assuming that ``single'' F~stars 
with $\Delta M_{c_0} > 0.15$ are in fact tight binary pairs, the fraction of 
such binaries is at least $\sim 40$~\%. Thus, close binaries may heavily impact 
stellar population statistics used to probe the formation and evolution of 
the Galaxy. Therefore, these stars are important not only for better 
understanding of stellar evolution, but also for developing 
the adequate picture of the history of the Galaxy.

In conclusion, we can  summarize our results as follows.
(i). The unresolved binary F~stars from the {\it HIPPARCOS} are typically much
brighter than predicted on the basis of their \uvby colors.
(ii). This discrepancy, called here brightness anomaly, inversely correlates 
with the binary components separations, suggesting that at least part of 
it is associated with binary components interaction in tight pairs. 
(iii). Brightness anomaly 
correlates with stellar kinematics in a way that implies higher, on average,
anomaly values at older ages. (iv). With ages from the same set of isochrones,
the age--velocity relations of the normal single stars and binary candidates 
with brightness anomaly (C~binaries) are discrepant in a way implying that the 
actual ages of C~binaries are older than predicted. 

These results argue that the main
sequence evolution of the primary F~star in a tight binary pair is different 
from that of a similar single star, evidently because of the interaction with
the secondary. The difference appears to be consistent with extra mixing in the
convective  core of the primary, presumably induced by that interaction; 
however, other mechanisms need to be carefully studied 
before a definitive conclusion can be reached.

\begin{acknowledgement}
I would like to thank the anonymous referee for helpful comments and 
suggestions. It is my pleasure to thank Claus Leitherer, Stefano Casertano,
and Nikolai Piskunov for stimulating discussions.
I am grateful to S.~Lyapustina and G.~Galas for careful reading the 
paper and numerous revision suggestions.
\end{acknowledgement}

{}


\begin{thebibliography}{}

\bibitem{}
Andersen, J. \&\ Nordstr\"om, B. 
1999, in: Theory and Tests of Convection in Stellar Structure (TTCSS),
ASP Conf. Ser., 173, 31
 
 \bibitem{}
 Carlberg, R.G., Dawson, P.C., Hsu, T., \&\ VandenBerg, D. 1985, ApJ, 294, 674

\bibitem{}
Chiosi, C. 
1999, in: TTCSS,
ASP Conf. Ser., 173, 9
 
\bibitem{}
Clausen, J. V., Baraffe, I., Claret, A., \&\ VandenBerg, D. A. 
1999, in: TTCSS,
ASP Conf. Ser., 173, 265
 
\bibitem{}
Crawford, D.L. 1975, AJ 80, 955

\bibitem{}
Demarque, P., Chaboyer, B., Guenther, D., Pinsonneault, M., 
Pinsonneault, L., \&\ Yi, S. 
1996, Yale Isochrones 1996 in ``Sukyoung Yi's WWW Homepage''
 
\bibitem{}
Gimenez, A., Claret, A., Ribas, I., \&\ Jordi, C. 1999,
in: TTCSS,
ASP Conf. Ser., 173, 41

\bibitem{}
Evans, D. W., van Leeuwen, F., Penston, M. J., Ramamani, N., 
\&\ Hog, E.  1992, A\&A, 258, 149

\bibitem{}
Griffin, R. F. \&\ Suchkov, A. A. 2001, in preparation

\bibitem{}
Hauck, B. \&\ Mermilliod, M. 1998, A\&AS, 129, 431

\bibitem{}
Mignard, F., Froschle, M., Badiali, M., Cardini, D., Emanuele, A., 
Falin, J. L., \&\ Kovalevsky, J.  1992,  A\&A, 258, 165

\bibitem{}
Mignard, F., Froschle, M.,  \&\ Falin, J. L.  1992,  A\&A, 258, 142

\bibitem{}
 Moon, T. T.  1985, Commun. Univ. London Obs., No. 98


\bibitem{}
Moon, T.T. \&\ Dworetsky, M.M. 1985, MNRAS, 217, 305
 
\bibitem{}
Pols, O. R. \&\ Marinus, M. 1994, A\&A, 288, 475

\bibitem{}
Schuster, W. J. \&\ Nissen, P. E. 1989, A\&A, 221, 65

\bibitem{}
S\"oderhjelm, S., Evans, D. W., van Leeuwen, F., \&\ Lindegren, L. 1992, 
A\&A, 258, 157

\bibitem{}
Suchkov, A. A. 2000, ApJ, 535, L107
 
\bibitem{}
Suchkov, A. A. \&\ McMaster, M. 1999, ApJ, 524, L99
 
\end{thebibliography}
\end{document}